# Dual-camera high-speed imaging of *n*-hexane oxidation in a high-pressure shock tube

**Miguel Figueroa-Labastida, Touqeer Anwar Kashif, Aamir Farooq***

King Abdullah University of Science and Technology (KAUST), Clean Combustion Research Center, Physical Sciences and Engineering Division, Thuwal 23955-6900, Saudi Arabia
*Corresponding author email: aamir.farooq@kaust.edu.sa

**Abstract**

Shock tubes are widely used in the study of chemical kinetics. Its benefits rely on the almost ideal shock-heating process that provides high temperatures and pressures to a chemical system for a limited test time. Just like any reactor, shock tubes are not immune to non-ideal effects. The study of conditions that might deviate experiments from ideal conditions is thus of the utmost importance. High-speed imaging has been proven to be a powerful bytool to analyze non-ideal / non-homogenous combustion in shock tubes. In this work, dual-camera high-speed imaging experiments were performed at 10, 15 and 20 bar in a high-pressure shock tube (HPST). An optical section was designed as an extension of the HPST which enabled simultaneous visualization from the endwall and the sidewall of the driven section of the shock tube. *n*-Hexane, a fuel with a negative temperature coefficient (NTC) behavior that has been identified as prone to non-homogenous ignition, is used as a test fuel. Reactive mixtures and thermodynamic conditions were selected to visually analyze ignition processes at the high-temperature, NTC and low-temperature regimes. Non-homogeneous ignition was observed mostly at the local maximum of the IDT, which is comprised by the high-temperature and NTC regions. Stoichiometric n-hexane mixture with high fuel loading (5% *n*-hexane) presented the highest deviation from constant volume chemical kinetic simulations. The inclusion of helium as a bath gas to mitigate preignition was tested and it showed to improve the susceptibility of the mixtures to develop reaction fronts. The modified Sankaran criterion for the identification of ignition regimes in shock tubes was tested and it showed an overall good agreement against the experimental observations.

## 1. Introduction

The goal of many chemical kinetic studies is the development and validation of models that enable understanding and optimization of complex chemical systems. Such studies involve measurements of key characteristic quantities, e.g., reaction rate coefficients, product yields, species time-histories, extinction rates, ignition delay times and flame speeds. These experiments are usually performed in chemical reactors such as flow reactors, jet-stirred reactors, shock tubes and rapid compression machines. Shock tubes have been extensively used for the study of fuel pyrolysis [1], oxidation [2] and reaction rates [3,4], and are usually considered as ideal homogeneous reactors allowing the study of the chemical evolution of a system at the desired temperature and pressure. Non-idealities in shock tube experiments jeopardize the validity of these assumptions and the quality of the acquired data. Therefore, any unwanted phenomenon occurring during the experiments needs to be identified, quantified and mitigated as much as possible.

Shock tube non-idealities can stem from a variety of sources and include gasdynamic effects [5–11] and the influence of diaphragm particles [12–14]. Gasdynamic effects are related to the attenuation of the incident shock wave [9], boundary layer separation, shock bifurcation [7,8], and gradual pressure rise (dP/dt) behind reflected shock [15,16]. Leftover diaphragm particles, which remained in the system even after cleaning, have been observed and analyzed during combustion processes and their influence has been assessed [12,13].

Preignition, described as weak ignition or non-homogeneous ignition, has been observed in several chemical systems [13,17–31]. Figueroa-Labastida et al. [19] correlated the proneness to preignition with the existence of temperature inhomogeneities in the system and thermochemical properties of the reactive mixture. The use of high-speed imaging has been key in understanding how this phenomenon unfolds and in identifying the regimes of conditions where non-homogeneous ignition is observed [12,13,17–19,22,25,26,29,32–36]. Due to the limitations in optical access and the strength of materials used in imaging experiments, researchers often rely on visualizing combustion at relatively low pressures. Therefore, experimental works using imaging to characterize combustion phenomena in high pressure shock tubes are scarce in literature. Using rectangular shock tubes, Fieweger et al. [29] reported the use of schlieren imaging of *iso*-octane and *n*-heptane at pressures of 13.6 and 14 bar, respectively, while Merkel and Ciccareli [34] analyzed methane combustion at a nominal pressure of 10 bar. Using circular shock tubes, only Lee et al. [25] reached a pressure of 10 bar during lateral schlieren visualization of ethanol combustion. Recently, with the introduction of sapphire endwalls, Nativel et al. [37] visualized ethanol ignition at pressures around 20 bar utilizing two high-speed cameras for of OH* and natural luminosity imaging. They identified Ar-diluted mixtures to be more prone to non-homogeneous ignition and were able to discern particle-driven ignition and local gas ignition. Shao et al. [38] studied the combustion of *n*-heptane at pressures near 15 bar and identified higher fuel loading promoting non-homogeneous ignition.

Previous work from Figueroa-Labastida et al. [17] characterized the non-homogeneous ignition of *n*-hexane in dual camera experiments using the Low-Pressure Shock Tube facility at KAUST, where the limitation on test times due to the reachable pressures in the facility did not allow the exploration of the NTC (negative temperature coefficient) region. In this work, simultaneous sidewall and endwall imaging experiments of *n*-hexane are reported at high pressures for the first time with a focus on the ignition processes observed in the NTC region. These experiments are intended to shed light into the behavior of fuel mixtures at engine-relevant conditions, and the existence and characteristics of preignition phenomena at high pressures.

## 2. Experimental details

A new end section that provides optical access from the sidewall and endwall was designed and manufactured, as depicted in Fig. 1. This section was attached to the driven section of the to High-Pressure Shock Tube (HPST) facility at KAUST. This section consists of a stainless-steel tube with two 140 mm long sidewall window ports. In contrast with the optical section in the Low-Pressure Shock Tube [32] consisting of two parallel window slits, the longitudinal windows in the HPST are situated 90 degrees from each other. This is intended for future applications of laser scattering techniques for further characterization of the combustion processes. The sapphire slit windows have a total thickness of 33 mm, a flat inner surface and a step-like design. Width of the windows is 20.5 mm and this width was selected to minimize the mismatch of the curvature of the stainless-steel body to the window surface. A transparent endwall was designed to enable visualization of the entire circular section of 101 mm diameter. A stepped configuration was selected so the inner surface of the endwall would match the edge of the slit window, leaving no blind spots in the test section. This allows a complete longitudinal visualization of the ignition processes from the endwall to a distance of 140 mm. Total thickness of the endwall window is 158 mm. This optical test section is designed for a maximum pressure of 100 bar.

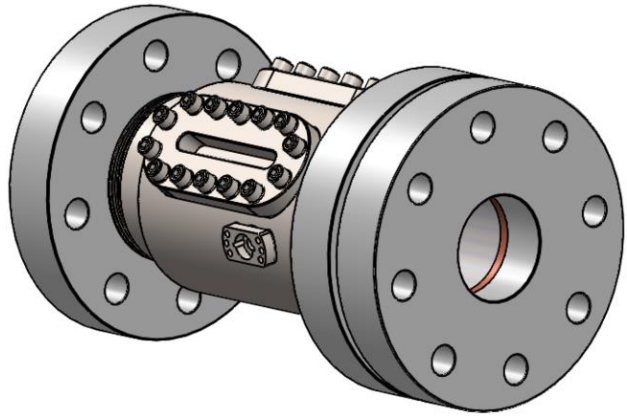

Figure 1. CAD drawing of the new optical section of the KAUST HPST.

Simultaneous lateral and endwall natural luminosity from autoignition experiments was recorded using two high-speed Photron Fastcam SA-X2 cameras (see Fig. 2), which were synchronized and images were recorded at 100,000 frames per second. Pressure rise of the incident shock wave was used as a trigger for recording the camera signals. UV lenses with optimized transmissivity between 250-650 nm and a focal length of 105 mm were used for both cameras. Only one slit window was used in the measurements reported herein. The test section was cleared of any debris after every run and the full length of the shock tube was cleaned whenever a new mixture was tested. Experiments were repeated by replacing the sidewall window with a blank and no significant differences were observed on the measured ignition delays or the two-dimensional qualitative behavior of ignition observed from the endwall window.

The mixtures analyzed in this study are detailed in Table 1. Compositions were selected based on previous studies [17,39] and pressures were chosen to observe similar total ignition delay times regardless of fuel loading.

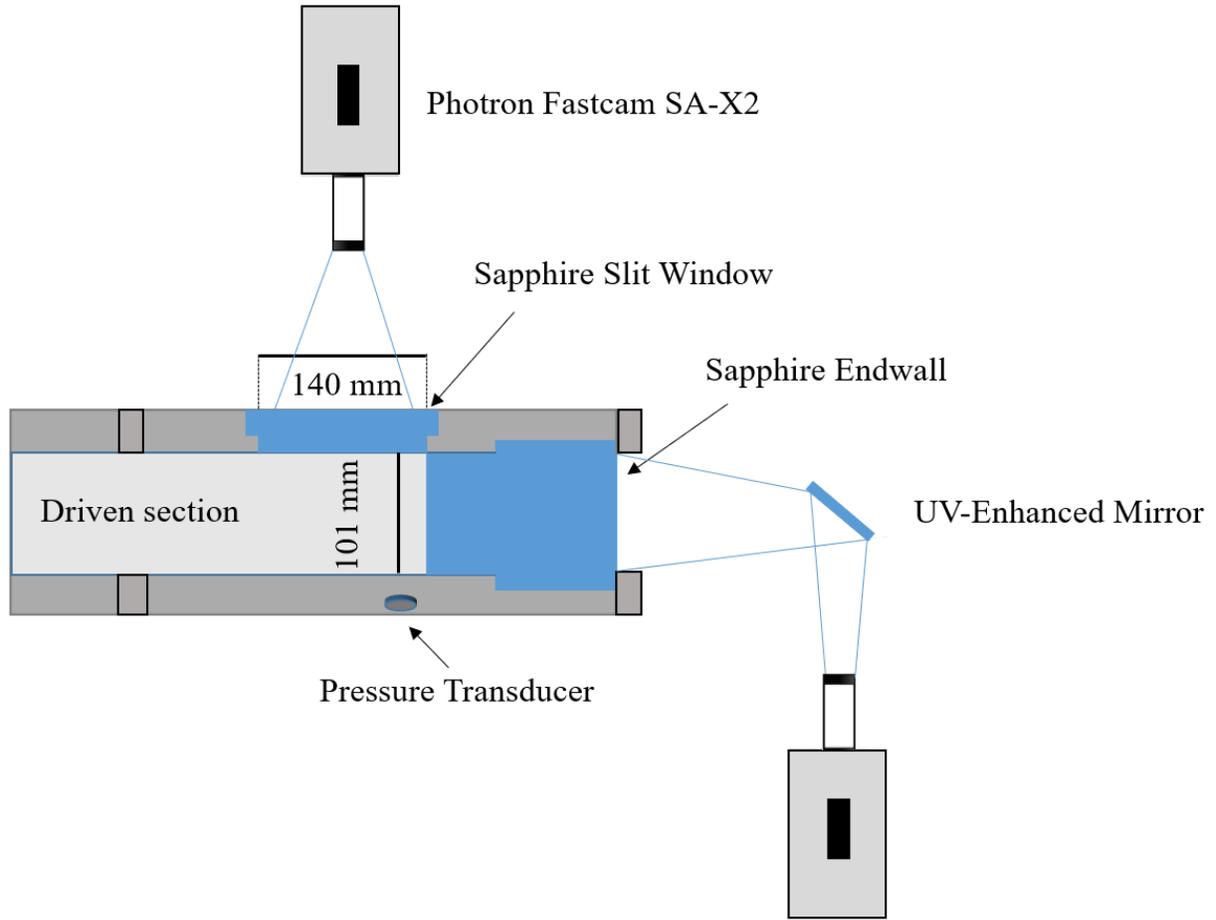

Figure 2. Experimental setup for simultaneous imaging experiments in the KAUST high-pressure shock tube.

Table 1. Compositions of the studied fuel mixtures.

| Fuel | Mixture | $X_{Fuel}$ [%] | $X_{O2}$ [%] | $X_{Ar}$ [%] | $X_{N2}$ [%] | $X_{He}$ [%] | $\phi$ | $P_5$ [bar] |
|---|---|---|---|---|---|---|---|---|
| *n*-hexane | H1 | 2.2 | 20.5 | 0 | 77.3 | 0 | 1 | 15 |
| | H2 | 5 | 23.75 | 71.25 | 0 | 0 | 2 | 15 |
| | H3 | 5 | 47.5 | 47.5 | 0 | 0 | 1 | 10 |
| | H4 | 5 | 47.5 | 0 | 0 | 47.5 | 1 | 10 |
| | H5 | 2.5 | 23.75 | 73.75 | 0 | 0 | 1 | 20 |

## 3. Results and discussion
### 3.1. Stoichiometric mixture H1 at 15 bar

Firstly, a stoichiometric mixture of n-hexane was tested at 15 bar. This mixture was studied by Zhang et al. [39] and their results are used to validate the measurements carried out here with the optical section. Figure 3 shows the measured ignition delay times of mixture H1. A good agreement is observed with results reported in the literature. Both measurements seem to be overpredicted by the chemical kinetic model of Sarathy et al. [40] at the onset (900 – 1000 K) of the NTC region. A dP/dt = 3%/ms simulation is not able to capture the deviation observed in the experimental measurements.

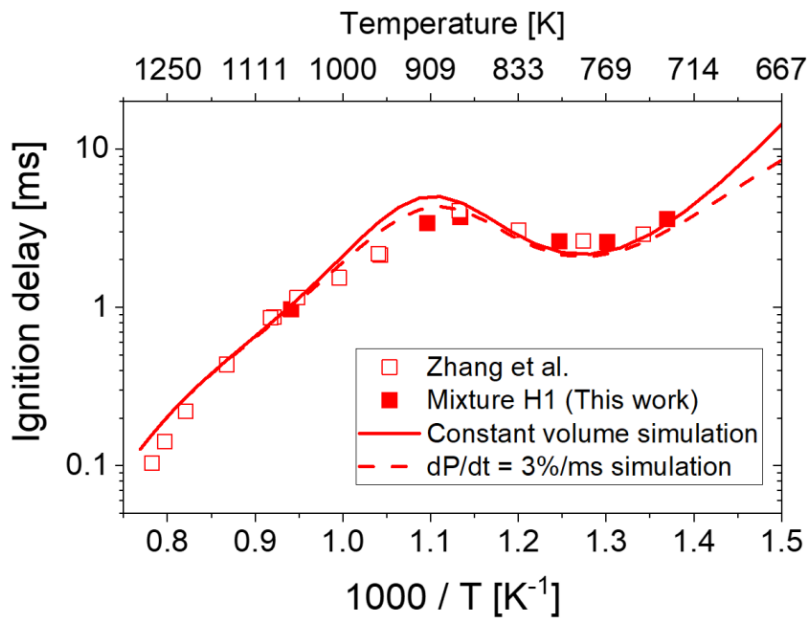

Figure 3. Ignition delay times of mixture H1. Experiments from Zhang et al. [39] and simulations using the model of Sarathy et al. [40] are used for comparison.

Images recorded during these experiments are depicted in Fig. 4. Rows are individual experiments with decreasing temperatures from top to bottom. Endwall images show the circular cross section of the shock tube, while the endwall is located at the left in the sidewall images. The distortion observed at top left of the endwall image is an optical imperfection in the endwall window which did not interfere with the test section. At high temperatures (1062 K), a nearly ideal combustion process is identified, where the mixture is consumed homogeneously in ~60 microseconds and the ignition event starts at the endwall, as expected. The experiment at 912 K, near the maximum of Fig. 3, shows a significantly different behavior. Here, the combustion process is seen to start from the left side of the shock tube cross section and develops in a slower manner, with the first ignition feature occuring ~80 mm away from the endwall. Entering the NTC region (801 K case), a fast and homogeneous ignition process is identified with ignition initiating at the endwall. In the low-temperature region (730 K case), the endwall image shows a non-homogeneous process and the ignition initiating ~30 mm away from the endwall.

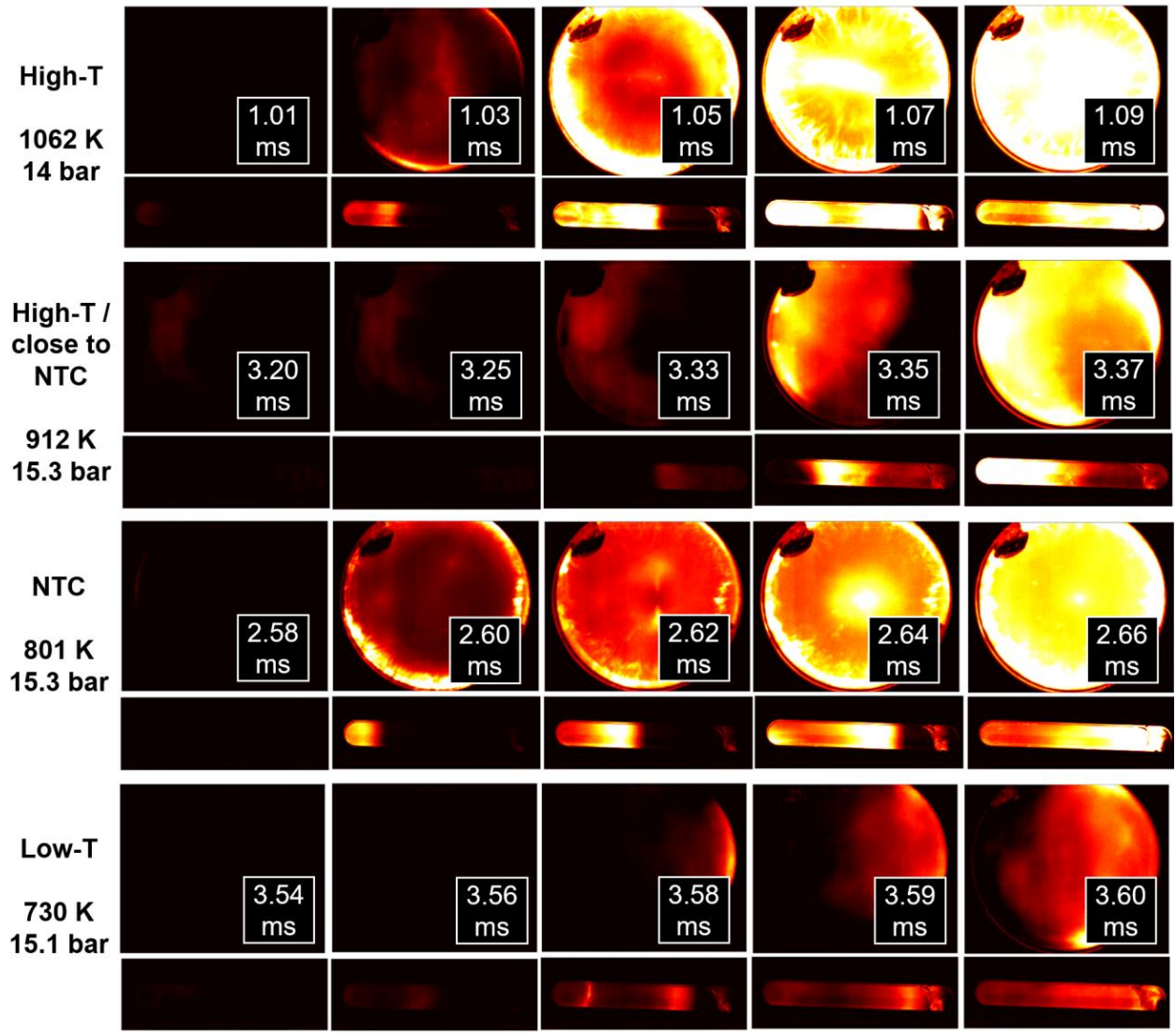

Figure 4. High-speed images of the combustion of mixture H1 at various temperatures and pressure of ~ 15 bar.

Interestingly, the NTC behavior of *n*-hexane allows for the observation of different temperature regimes where non-homogeneous ignition is dominant, in contrast with non-NTC fuels like methanol or ethanol [19], with which these observations were exclusive of the lower temperature regime. The weak ignition case at 912 K, observed through high-speed imaging as a slow and far-wall ignition, correlates with the observation of a shorter ignition delay time compared to the value predicted by the kinetic models; while the case at 730 K, in the lower temperature regime, does not show an expedition of the delay time with respect to the model prediction. One aspect to notice is that the time from the first observation of emission until ignition in the 912 K case is much longer than the 730 experiment, thus showing a higher severity of preignition, related to a characteristic residence time for non-homogeneities in the test section ($\tau_{nh)}$ [17] . The strong ignition cases, at 1062 K and 801 K, do not show faster delays than the models' simulations as expected.

### 3.2. Rich mixture H2 at 15 bar

Ignition delay times of the fuel-rich mixture H2, containing argon as the bath gas, are shown in Fig. 5. The measured values agree well with Sarathy et al. [40] model with some underprediction in the valley of the NTC region. The fuel-rich mixture ($\varphi = 2$ in air) investigated by Zhang et al. [39] is plotted for comparison with constant volume simulations using the same model [40]. In this case, there is also a similar underprediction of their experimental values in the NTC region.

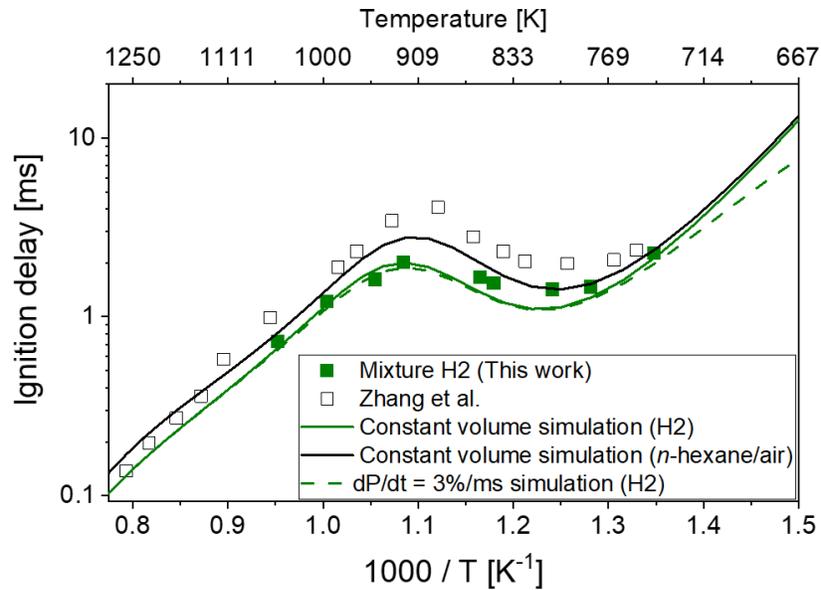

Figure 5. Ignition delay times of mixture H2. Measurements of a rich mixture ($\phi = 2$) of n-hexane in air from Zhang et al [39] are plotted for comparison. Constant volume simulations were performed using the model by Sarathy et al. [40]

Figure 6 showcases the combustion processes for this mixture across different temperature regimes. At high temperatures (e.g., 1049 K case), we observe a fast ignition happening near the endwall. Nevertheless, the process is not fully homogeneous, with an initially localized emission front that expands until the entire cross section is ignited. In the NTC region (858 K case), a similar process is observed with a different radial location of the starting ignition feature. Another example in this region (805 K case) behaves similarly to the previous cases with the starting ignition occurring from the top center of the tube. In the low-temperature region (741 K case), the ignition process appears to be similar to that of the high-temperature example (1049 K case) with the main difference being the time (0.73 ms vs 2.45 ms) when emission starts to be identified. No significant deviation from the kinetic models is observed despite the localized nature of the ignition fronts.

### 3.3. Stoichiometric mixtures H3 and H4 at 10 bar (influence of He)

Ignition delay times of the stoichiometric mixture H3 (5% n-hexane, 47.5% oxygen in Ar) are plotted in Fig. 7. This mixture has relatively low level of dilution (47.5% bath gas) and is expected to be more prone to non-homogenous ignition. A clear deviation between the measured delay times and the constant volume simulation is observed in the high-temperature and NTC regions. At lower temperatures, the experiments and model predictions have a reasonable agreement.

The use of helium bath gas as a mitigating agent of non-homogeneous processes in shock tubes was suggested by Nativel et al. [31], who observed that adding 10% He could eliminate preignition and the expedited ignition delays. To test their hypothesis, we performed experiments with mixture H4, where the Ar content (47.5%) of H3 was fully substituted with He. IDTs of mixture H4 are also plotted in Fig. 7. Some minor improvement is observed near the top of the high-temperature region and in the NTC region, but the measured ignition delays still appear to be too short compared to the both the constant volume and dP/dt = 3% model predictions.

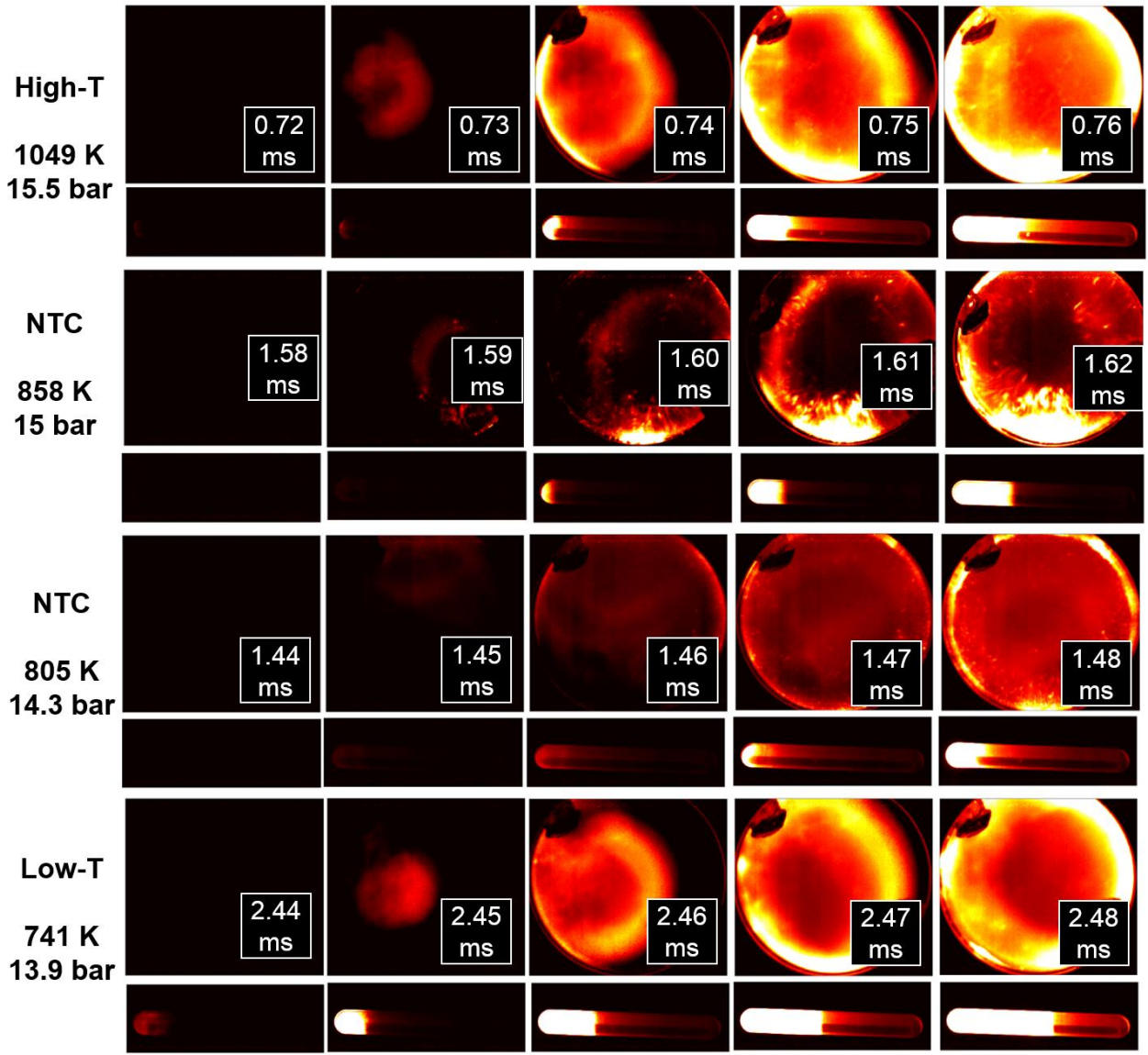

Figure 6. High-speed images of the combustion of mixture H2 at various temperatures and pressure of ~ 15 bar.

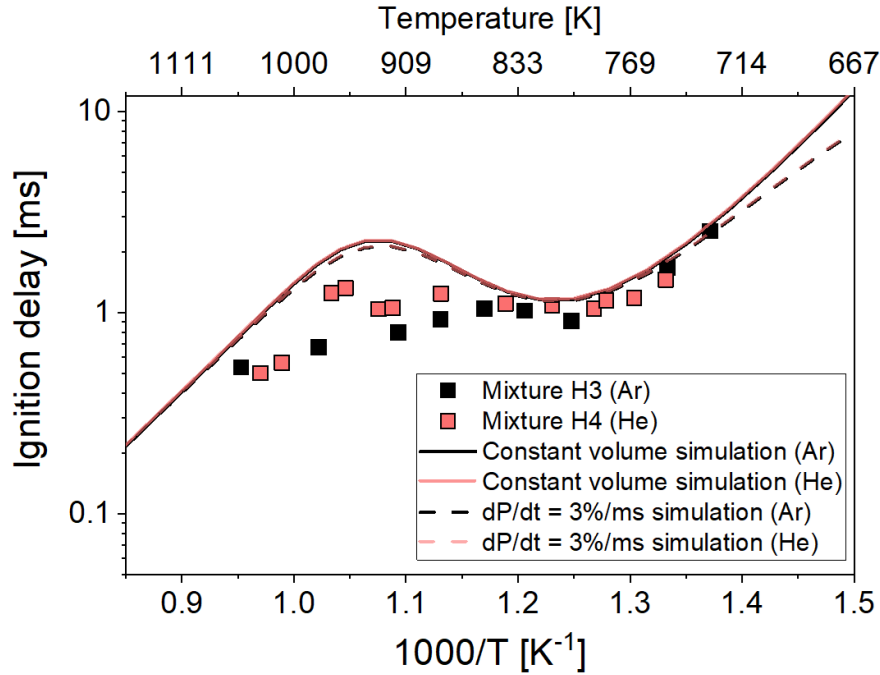

Figure 7. Ignition delay times of mixture H3 and H4 at 10 bar. Constant volume simulations were performed using the model by Sarathy et al. [40] . An overlap of the simulation predictions is observed.

Figure 8 shows a comparison of the ignition of mixtures H3 and H4 using high-speed imaging. The first two rows compare high-temperature points (967 K, 978 K) where a significant difference was observed in the ignition delay times of the two mixtures. The He-containing case is characterized by a very fast radial process, occurring in ~20 microseconds, with the location of the ignition being ~35 mm from the endwall. The Ar-containing case, on the other hand, shows early appearance of an ignition kernel which propagates from the left side of the cross section and a second slow feature appearing from the right until a fast front is observed from the bottom right. In the NTC region (884 K cases), where ignition delays of H3 and H4 were similar, non-homogeneous processes are observed in both mixtures. The He-containing mixture shows ignition from the bottom right of the shock tube cross section, while the Ar-containing mixture shows an earlier emission feature from the right side of the cross section with both cases igniting near the endwall. The inclusion of He changes the phenomenological behavior of the reactive system in the presence of preignition sources, demonstrating a higher resistance to the formation of early ignition kernels. Nevertheless, the sole inclusion of He does not completely mitigate non-homogeneous ignition at the conditions studied in this work and a significant deviation from the delay times predicted by chemical kinetic models is remanent.

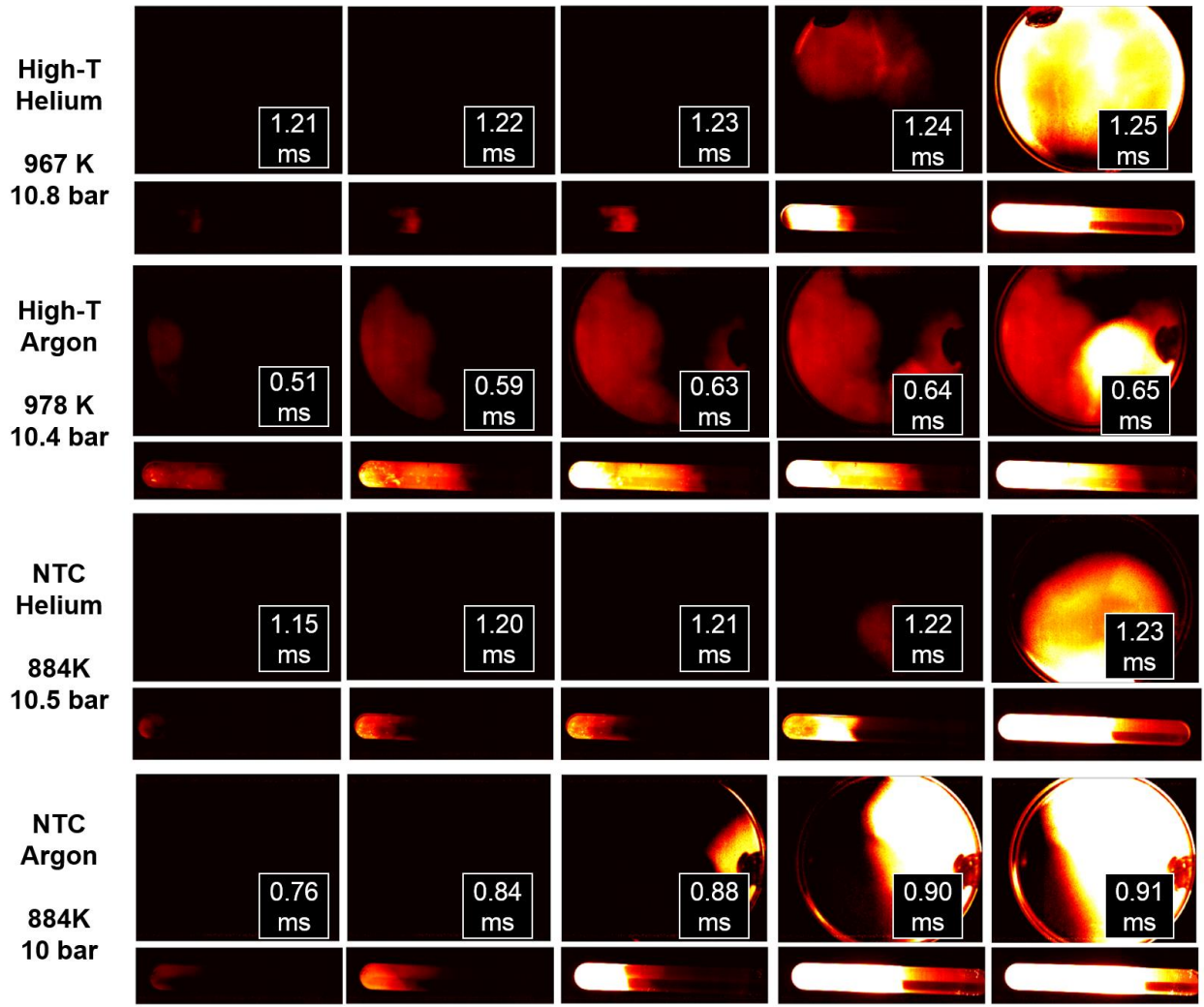

Figure 8. High-speed images of the combustion of mixture H3 and H4 at two temperatures and pressures ~ 10 bar.

### 3.4. Stoichiometric mixture H5 at 20 bar

In order to explore the effect of pressure on the ignition characteristics of *n*-hexane, a stoichiometric mixture (2.5% n-hexane) was tested at 20 bar. Ignition delay times of mixture H5 are plotted in Fig. 9 and compared with constant volume simulations. Ignition delays are shorter than the model prediction at high temperatures and near the peak of the NTC region, while the model slightly overpredicts measured ignition delays in the valley of the NTC region.

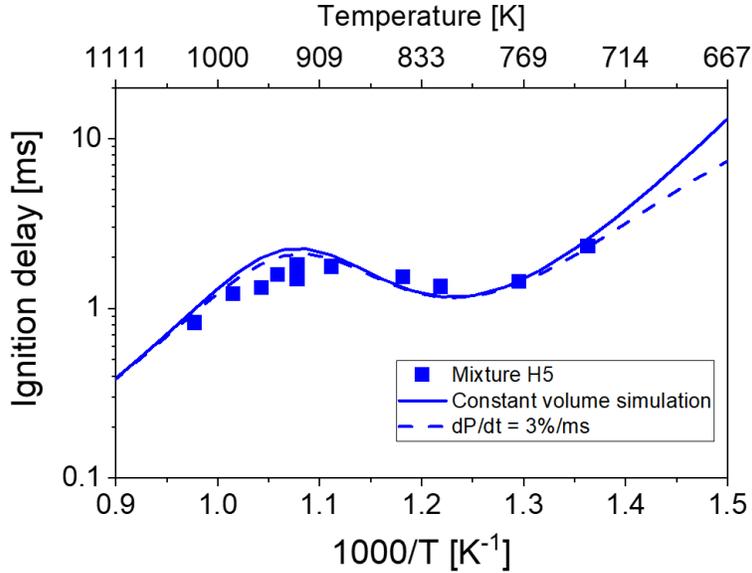

Figure 9. Ignition delay times of mixture H5 at 20 bar. Constant volume simulations were performed using the model of Sarathy et al. [40].

Figure 10 depicts high-speed images of the combustion processes of H5 at different temperature regimes. In the high-temperature region (985 K case), where the measurements deviated significantly from the model, a non-homogeneous ignition process is observed. An initial emission kernel is identified at the bottom of the shock tube cross section which then expands along the edges. A second front, which can be clearly identified through the sidewall imaging, is formed far from the endwall. A third strong front is observed to originate from the left edge and it completes the ignition process in ~240 microseconds. At lower temperature (927 K case), close to the local maximum in the ignition delay curve, multiple ignition features occurring away from the endwall are observed. Radial ignition starts from the bottom and the front travels along the edges, consuming the mixture close to the walls before igniting the core of the cross section. In the NTC region (900 K case), ignition first occurs at the endwall with a second front appearing ~125 mm away from the endwall. Ignition occurs first in the right side of the shock tube cross section and expands along the edges until a second fast feature appears from the bottom side, consuming the mixture quickly. In the low-T region (770 K case), a fast process is observed at the endwall which takes ~ 40 μs to consume the entire mixture. In summary, at the higher pressure of 20 bar albeit with lower fuel loading, far-wall ignition is observed even at high temperatures and localized ignition from the bottom of the shock tube cross section seems to be favored for most of the cases.

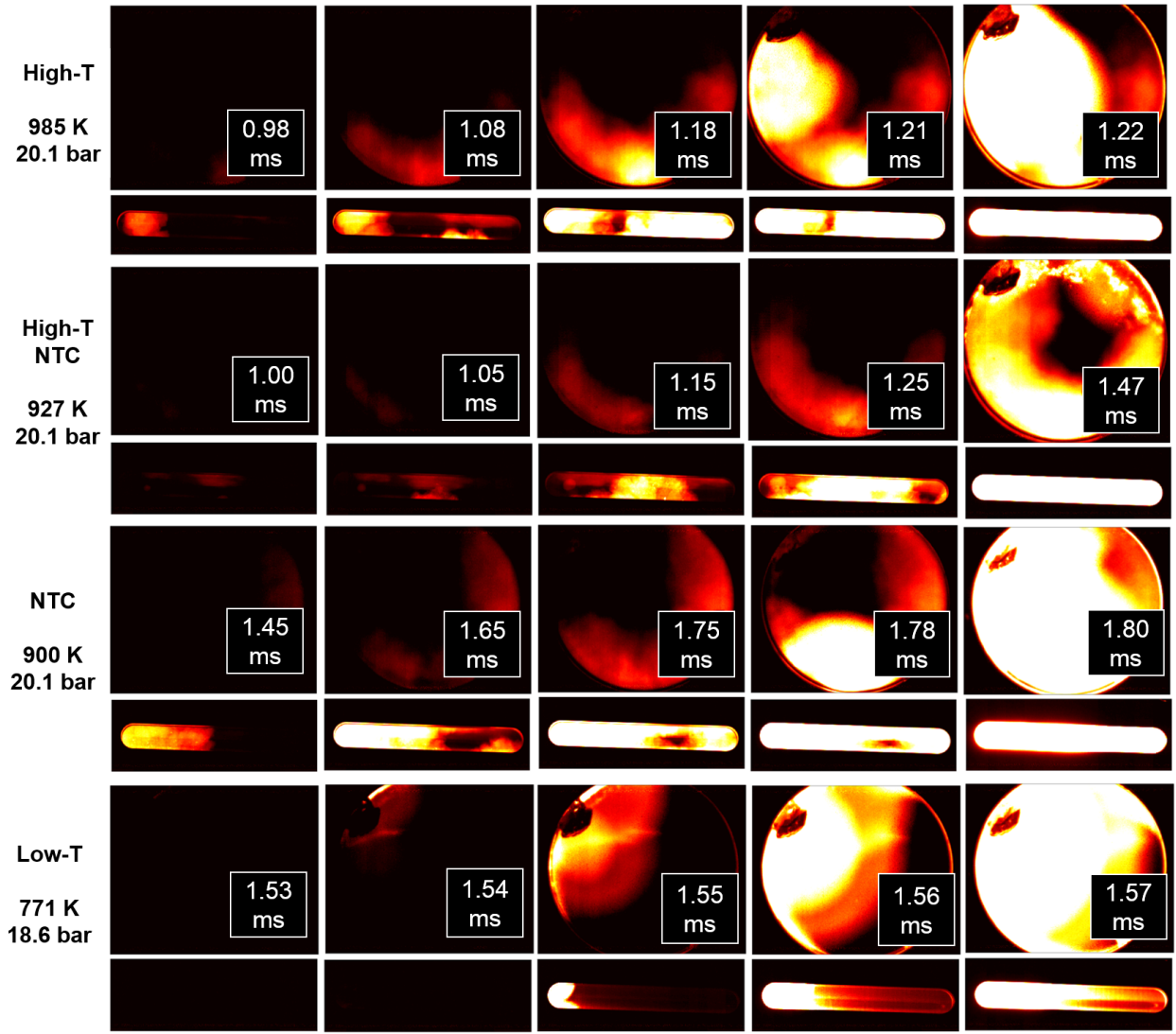

Figure 10. High-speed images of the combustion of mixture H5 at various temperatures and pressure ~ 20 bar

## 4. Identification of ignition regimes

Previous works mostly observed non-homogeneous ignition at low temperatures [13,17–31]. In this work, irregular ignition events are observed at almost all temperature regions (high, NTC, low). In order to shed light on the thermochemical properties of the fuel/oxidizer mixture that may promote the likelihood of non-homogeneous ignition at various temperature regions, a criterion for ignition regimes needs to be explored and tested against the experimental observations.

The modified Sankaran criterion [19] was introduced for the analysis of the homogeneous / non-homogeneous ignition of various mixtures of ethanol and methanol in shock tube, and was successful in predicting the regimes identified by high-speed imaging. The modified Sankaran number $Sa_p$ is defined as:

$$Sa_p = \beta S_L \left|\frac{d\tau_{ig,i}}{dT}\right| \frac{T'}{l_T Re_T^{-0.5}}$$

where $\beta=0.5$ is a constant, $d\tau_{ig,i}/dT$ is the ignition delay sensitivity of the mixture computed from constant volume simulations (Sarathy et al. [40] model), $T'$ is the root mean square (RMS) temperature variation of the core gas (considered here to be 2 K), $l_T$ is the characteristic length scale of the temperature fluctuation field (estimated to be 1 mm [13]), and $Re_T$ is defined as $Re_T = T' l_T / \alpha$ where $\alpha$ is the thermal diffusivity of the mixture. Further description of this criterion can be found in [19]. It may be noted that the calculation of $Sa_p$ is quite sensitive to the values of $T'$ and $l_T$, and the values used here are the best estimates in the absence of any precise determinations.

Modes of ignition of a specific mixture can be classified using $Sa_p = 1$ as the threshold value that separate the weak / non-homogeneous ($Sa_p > 1$) and strong / homogeneous ($Sa_p < 1$) ignition regimes. $Sa_p$ is plotted in Fig. 11a for mixture H1 at 15 bar. The shape of the $Sa_p$ line differs drastically to those in our previous work [19] due to the existence of an NTC region and its effect on the ignition delay sensitivity across the temperature range. Here, it is seen that the threshold value of $Sa_p = 1$ is crossed at two points. Firstly, at the region between 940 – 983 K, where the maximum value of $Sa_p = 1.14$ occurs at 958 K. Referring to the IDT plot in Fig. 3, it is clear that this region showed the highest deviation between the model predictions and the experiments. Expedited ignition is the typical characteristic of the existence of non-homogeneous ignition. In Fig. 4, the only case that portrays a clear non-homogeneous and remote ignition occurs at 912 K, which is close to this region but has a predicted value of $Sa_p = 0.3$. Secondly, the threshold is crossed at temperatures lower than 727 K. Although the case at 730 K does not show much deviation from the model IDT simulation of Fig. 3, the images shown in Fig. 4 evidence localized ignition occurring at two axial points far from the endwall.

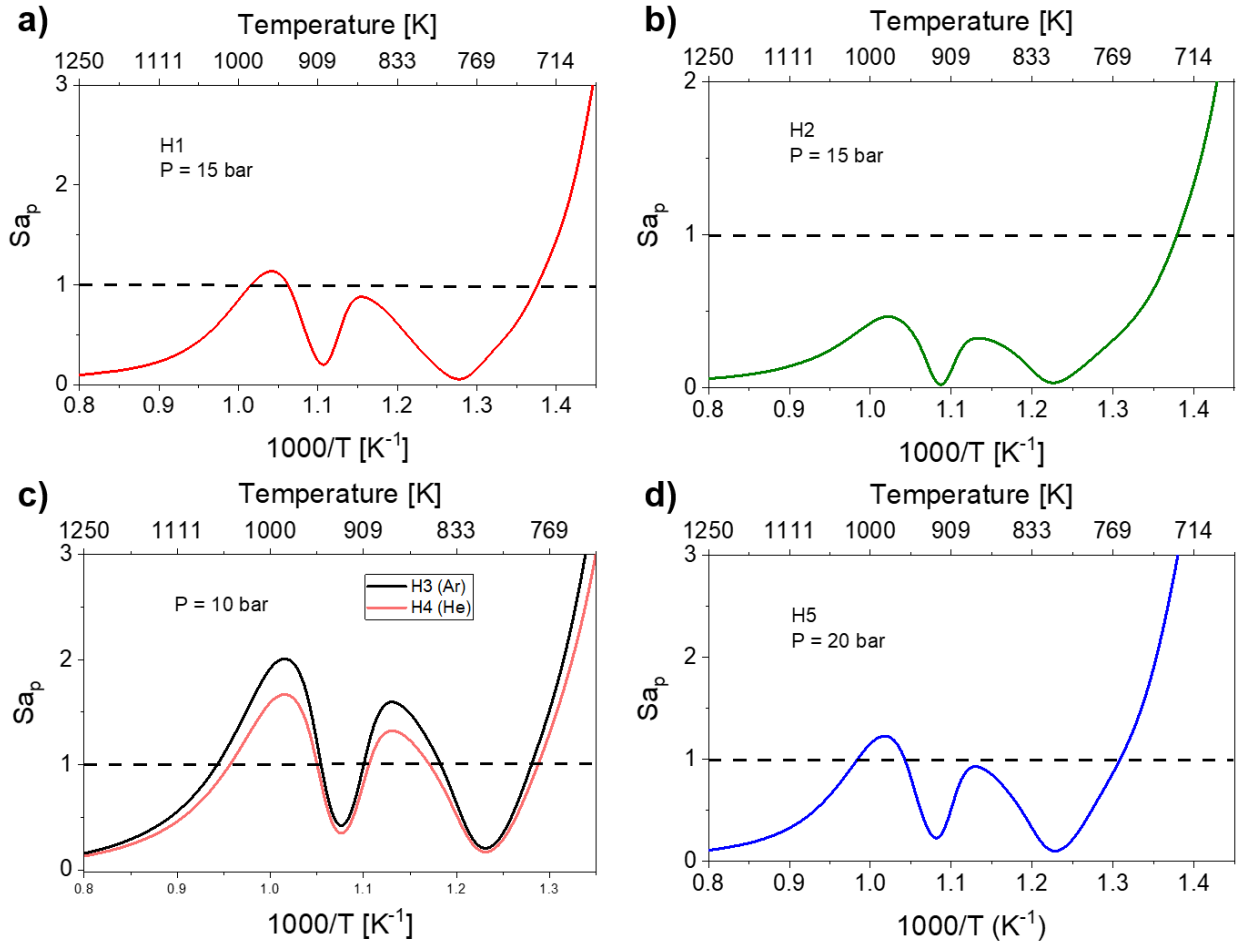

Figure 11. Computed $Sa_p$ values for mixtures H1 – H5.

Figure 11b shows the regime prediction of the rich mixture H2 at 15 bar. In contrast with H1, the $Sa_p = 1$ threshold is only crossed at low temperatures, below 725 K. All data points plotted in Fig. 5 are at higher temperatures than 725 K, and, therefore, non-homogeneous ignition would not be expected. In Fig. 6, the lowest temperature case at 741 K shows a relatively ideal ignition process, starting from the endwall similar to other depicted cases, but this case also has an initially localized but fast emission front, distinctive of a strong ignition.

The $Sa_p$ numbers for the high-fuel loading mixtures, H3 and H4, are plotted in Fig. 11c. The relative values of $Sa_p$ for the Ar-containing mixture are higher than those for the He-containing mixtures.. This shows that an improvement of the ignition strength is expected by He as the bath gas. For both mixtures, there are three different temperature ranges where the $Sa_p$ threshold is passed. For H3, these are regions 948-1060 K, 845-908 K and below 781 K. This agrees well with the trends of Fig. 7, where a clear and large deviation is observed from the cases over 854-1049 K. The only case outside this range at 914 K still shows a large deviation from the predicted values. The cases at temperatures lower that 781 K do not seem to deviate from the model predictions. The images of Ar-containing case at 978 K (see Fig. 8), corresponding to an $Sa_p$ value close to the maximum ($Sa_p = 2$), show characteristics of a slow localized ignition. The case at 884 K, with a computed value of $Sa_p = 1.59$ and close to the second local maximum in Fig. 13, also depicts localized ignition in Fig. 8 but with lower residence time that the 978 K case. For He-containing mixture H4, the deviation from the IDT trend is lower but it still exists at the same temperature range described for H3. Here, the $Sa_p$ values are higher than 1 in the regions 953-1043 K, 854-903 K and below 776 K. The images from Fig. 8 at 967 K and 884 K, in the weak ignition regime, exhibit a localized but fast ignition front, with the ignition initiating far from the endwall for the case at 967 K.

Predicted $Sa_p$ values for H5 mixture at 20 bar are plotted in Fig. 11d. Here, similarly to mixture H1, there are two regions where a weak ignition regime is expected, 958-1014 K and below 764 K. The first region agrees well with the identified discrepancy between experiments and model predictions (see Fig. 9). However, the case at 771 K which is predicted to have an $Sp = 0.83$ did not exhibit a deviation from the predicted values. The highest temperature case shown in Fig. 10 at 985 K, corresponding to the local maximum of $Sa_p = 1.22$, shows ignition happening at multiple axial locations far from the endwall and a slow emission front which grows until a fast reaction front ignites the mixture. The cases at 927 K ($Sa_p = 0.25$) and 900 K ($Sa_p = 0.78$) exhibit similar non-ideal behavior with long residence times of localized ignition fronts and various simultaneous remote ignition positions.

The use of an *a priori* criterion such as the modified Sankaran number gives a good insight on the various ignition regimes, shedding light on how the thermochemical properties of the mixtures could impact the ignition phenomena at specific conditions of temperature and pressure. Here, it is shown that $Sa_p$ provides an adequate idea of how the NTC behavior of a fuel would affect the ignition regime. This is in contrast to the previous application of $Sa_p$ to non-NTC fuels, such as methanol and ethanol, where only low-temperatures cases were seen to be affected by non-homogeneous ignition. Although there are some specific conditions that are not predicted well by $Sa_p$ criterion here, the temperature fluctuation and characteristic length parameters, $T'$ and $l_T$, have a high impact on the computed values. Therefore, characterization of temperature gradients in various system is of utmost importance to improve the use of this and other predictive criteria.

## 5. Conclusions

Ignition behavior of various mixtures of *n*-hexane was analyzed using high-speed imaging in a high-pressure shock tube. Several phenomenological characteristics were observed at the high-temperature, NTC and low-temperature regimes of the ignition delay trends. Non-homogeneous ignition was observed systematically at temperatures marking the transition between the high-temperature and NTC regions, which occur at the local maximum of the ignition delay plots. Five mixtures varying in fuel-loading, equivalence ratio and bath gas were studied over a range of temperatures and pressures of 10-20 bar. Preignition development was observed with similar residence times of the reactive gases in the shock tube, with inhomogeneities seen as early as 0.51 ms in some cases. Mixture H1, with $N_2$ as bath gas, showed pronounced preignition close to the onset of NTC region at 15 bar, with far-wall ignition evident also at low-T. This in contrast with Ar-containing mixture H5, having a similar fuel loading at a higher pressure of 20 bar, where simultaneous localized ignitions were observed throughout the high-T and NTC regime. The effect of equivalence ratio was evident as the rich mixture H2 was identified to ignite locally with a nearly spherical propagation front with ignition happening close to the endwall. Other mixtures such as H5 tended to propagate along the edges of the shock tube. The effect of a lower dilution was depicted through mixtures H3, where slow emission and thus a long non-homogeneity residence time was identified. The inclusion of He as a preignition-mitigating bath gas was tested and it resulted in decreasing the susceptibility of the mixture to sustain and develop a reaction, but the effect was relatively minor. The modified Sankaran criterion for shock tubes was tested against the experimental observations. There was an overall good agreement on the regions where weak/non-homogeneous ignition was expected. Characterization of temperature fluctuations in shock tube facilities is identified as crucial to improve the predictions of this criterion.

**Acknowledgement**

The work reported in this publication was funded by King Abdullah University of Science and Technology (KAUST).